\documentstyle[12pt,fleqn]{article}
\def\singlespace {\smallskipamount=3.75pt plus1pt minus1pt
                  \medskipamount=7.5pt plus2pt minus2pt
                  \bigskipamount=15pt plus4pt minus4pt
                  \normalbaselineskip=15pt plus0pt minus0pt
                  \normallineskip=1pt
                  \normallineskiplimit=0pt
                  \jot=3.75pt
                  {\def\smallskip {\vskip\smallskipamount}}
                  {\def\medskip   {\vskip\medskipamount}}
                  {\def\bigskip   {\vskip\bigskipamount}}
                  {\setbox\strutbox=\hbox{\vrule
                    height10.5pt depth4.5pt width 0pt}}
                  \parskip 7.5pt
                  \normalbaselines}
\def\middlespace {\smallskipamount=5.625pt plus1.5pt minus1.5pt
                  \medskipamount=11.25pt plus3pt minus3pt
                  \bigskipamount=22.5pt plus6pt minus6pt
                  \normalbaselineskip=22.5pt plus0pt minus0pt
                  \normallineskip=1pt
                  \normallineskiplimit=0pt
                  \jot=5.625pt
                  {\def\smallskip {\vskip\smallskipamount}}
                  {\def\medskip   {\vskip\medskipamount}}
                  {\def\bigskip   {\vskip\bigskipamount}}
                  {\setbox\strutbox=\hbox{\vrule
                    height15.75pt depth6.75pt width 0pt}}
                  \parskip 11.25pt
                  \normalbaselines}
\def\doublespace {\smallskipamount=7.5pt plus2pt minus2pt
                  \medskipamount=15pt plus4pt minus4pt
                  \bigskipamount=30pt plus8pt minus8pt
                  \normalbaselineskip=30pt plus0pt minus0pt
                  \normallineskip=2pt
                  \normallineskiplimit=0pt
                  \jot=7.5pt
                  {\def\smallskip {\vskip\smallskipamount}}
                  {\def\medskip   {\vskip\medskipamount}}
                  {\def\bigskip   {\vskip\bigskipamount}}
                  {\setbox\strutbox=\hbox{\vrule
                    height21.0pt depth9.0pt width 0pt}}
                  \parskip 15.0pt
                  \normalbaselines}

\include{dspace12}
\def\be{\begin{equation}}
\def\ee{\end{equation}}
\def\bea{\begin{eqnarray}}
\def\eea{\end{eqnarray}}
\def\nn{\nonumber}

\def\lt{\left}
\def\rt{\right}
\def\sect #1{\setcounter{equation}{0}}

\begin{document}
\middlespace
\begin{center}
{\LARGE { Exact solutions of Einstein and Einstein-scalar equations in 2+1
dimensions}}
\end{center}
\vspace{1.0in}
\begin{center}
{\large{
K. S. Virbhadra\footnote[1]{E-mail\ :\ WTXVIXXK@LG.EHU.ES}\\
Departamento de F\'{\i}sica Te\'{o}rica\\
Universidad del Pa\'{\i}s Vasco\\
Apartado 644, 48080 Bilbao, Spain\\
}}
\end{center}
\vspace{1.5in}
\begin{abstract}
A nonstatic and circularly symmetric exact solution of the Einstein
equations (with a cosmological constant $\Lambda$ and null fluid) in
$2+1$ dimensions is given. This is a nonstatic generalization of the
uncharged spinless BTZ metric. For $\Lambda = 0 $, the spacetime is
though not flat, the Kretschmann invariant  vanishes. The  energy,
momentum, and power output for this metric  are
obtained. Further  a static and circularly symmetric exact solution
of the Einstein-massless scalar equations is given, which has a
curvature singularity at $r =0$ and the scalar field  diverges at
$r=0$ as well as at infinity .
\end{abstract}
\newpage
In recent years there is considerable interest in studying the Einstein
theory of gravity in $2+1$ spacetime dimensions. Due to the smaller number
of  dimensions it has tremendous mathematical simplicity.
The use of three dimensional gravity has been suggested as a test bed for the
quantization  of gravity $[1]$ . The lowest spacetime dimension in which the
Einstein theory of gravitation exists is three and it possesses some strange
features. In  three dimensions the number of independent components of the
Riemann curvature tensor and the Einstein tensor are the same (six) and  the
Riemann tensor can be expressed in terms of the Einstein tensor. Consequently,
the  spacetime described by the
vacuum solutions to the Einstein equations in $2+1$ dimensions are locally
flat and there are no gravitational  waves and no interaction between masses.
However, in   Newton's theory of gravitation in three dimensions a
point mass produces an acceleration which falls off like $1/r$ $[2]$.
Unlike the case in $3+1$ dimensions, the Einstein theory of gravitation has no
Newtonian limit in $2+1$ dimensions. Therefore, the Einstein
gravitational constant is undetermined in three dimensions.

Deser and
Mazur $[3]$ obtained the gravitational and Coulomb fields of a massive point
charge in $2+1$ dimensional Einstein-Maxwell (EM) theory.
As  opposed to the case of $3+1$ dimensional EM theory, the Coulomb
field behaves asymptotically as $1/r$ and the stress tensor  is nonvanishing
everywhere. The total energy diverges. The Kretschmann invariant
diverges at  $r =0$ as well as at infinity $[3]$. There exists  a static
multi-charge solution in $3+1$ dimensional EM  theory $[4]$. However, there
are no static solutions corresponding to two or more charges in $2+1$
dimensions $[3]$.
Arguing that the light emitted from the surface of a star will always
escape to the infinity because there are no gravity outside of matter (or
energy), Giddings, Abbott, and Kuchar concluded that black holes do not
exist in three dimensional Einstein theory of gravitation $[5]$. However,
Ba\~{n}ados, Teitelboim, and Zanelli (BTZ) $[6]$ have recently  discovered
a black hole solution to the  EM equations (with  a negative cosmological
constant) in $2+1$ dimensions, which is characterized by
mass, angular momentum, and charge parameters. The $2+1$ dimensional uncharged
BTZ black hole solution has fascinated many physicists minds ($[7]$ and
references  therein). Unlike the Kerr metric it is not asymptotically
flat and there is no curvature singularity. However, it shares many features
of its $3+1$ dimensional counterpart. For instance, it has an ergosphere,
upper bound in angular momentum ( for a given mass)  and has similar
thermodynamical properties $[8]$.

In this Letter we give two exact solutions : $(a)$ a nonstatic and circularly
symmetric solution of the Einstein equations with a cosmological constant and
null fluid, which is a nonstatic generalization to the spinless uncharged BTZ
metric and $(b)$ a static and circularly symmetric solution of the
Einstein-massless scalar equations. We obtain the energy, momentum, and power
output for the nonstatic circularly  symmetric metric (with zero
cosmological constant) and compare the results with its $3+1$ dimensional
counterpart. We use the geometrized units and follow the convention that
Latin (Greek) indices take values $0$ to $2$ ($1$ to $2$). $x^0$ stands for the
time coordinate.

The Einstein field equations with a cosmological constant
$\Lambda$ and the null fluid present are
\be
R_{ij} - \frac{1}{2} R\ g_{ij} + \Lambda\ g_{ij} =\  \kappa\  N_{ij}  \ .
\ee
$R_{ij}$ is the Ricci tensor and $R$ is the Ricci scalar. $\kappa$ is the
Einstein gravitational constant.
$N_{ij}$ is the energy-momentum tensor due to the null fluid, which is given by
\be
N_{ij} = V_i\ V_j \ ,
\ee
where $V_i$ is the null fluid current vector satisfying
\be
 V_i V^i\  =\ 0 \ .
\ee
A nonstatic and circularly symmetric exact solution of the above equations in
$2+1$ dimensions is given, in coordinates $ x^0 = u, x^1 = r, x^2 = \varphi$,
by the line element
\be
ds^2\ =\ - ( - m(u) - \Lambda\ r^2)\ du^2 - 2\ du\ dr + r^2\ d\varphi^2.
\ee
The nonvanishing component of the energy-momentum tensor of the null fluid
is
\be
N^1_0 \ = \ \frac{\dot{m}}{2\ \kappa\ r}\ \  .
\ee
The dot denotes the derivative with respect to the time coordinate $u$.
The null fluid current vector is
\be
V^i\ =\ g^i_1 \ \sqrt{\frac{- \dot{m}}{2  \kappa r}}\ \ .
\ee
The nonvanishing components of the Riemann curvature tensor are
\bea
R_{1220}\ &=&\ \Lambda\ r^2,  \nn\\
R_{1010}\ &=&\ - \Lambda,     \nn\\
R_{2020}\ &=&\ \frac{2 m \Lambda r^2 + r \lt( -\dot{m} + 2 \Lambda^2 r^3\rt) }
                  {2},
\eea
and the Kretschmann invariant is
\be
R^{abcd}\ R_{abcd}\ = \ 12 \Lambda^2 .
\ee
The spacetime described by the line element $(4)$
with  $\Lambda = 0$  is though not flat, the Kretschmann invariant vanishes.
When $m$ is a constant, $(4)$ gives the uncharged spinless BTZ metric.

Lindquist, Schwartz and Misner $[9]$ obtained the energy, momentum, and
power ouput for the Vaidya metric. It is of interest to obtain the same
for the circularly symmetric nonstatic metric ( with $\Lambda = 0$ )
and compare the results with its $3+1$ dimensional
counterpart. There are several prescriptions for calculating
the energy, momentum, and power output for a nonstatic general relativistic
system. For example, one can use the Landau and Lifshitz (LL) pseudotensor
$[10]$
\be
L^{mn} = L^{nm} = \frac{1}{2 \kappa} \ {S^{mjnk}}_{,jk}\ \ ,
\ee
where
\be
S^{mjnk}\ =\
 -g \left( g^{mn} g^{jk} - g^{mk} g^{nj}\right)\  .
\ee
$S^{mjnk}$ has the symmetries of the Riemann tensor.
The LL pseudotensor satisfies the local conservation laws:
\be
\frac{\partial L^{mn}}{\partial x^m}\ =\ 0 .
\ee
$L^{00}$ and $L^{0\alpha}$ are, respectively, the energy density and
momentum (energy current) density components. It is known that
the LL pseudotensor gives the correct result if
calculations are carried out in  ``Cartesian coordinates''.
Therefore, we transform the line element $(4)$, with $\Lambda = 0$, to
these coordinates, which is given by
\be
ds^2 = -dt^2 + dx^2 + dy^2 + \lt(1+m(u)\rt)\ \lt[
       dt\ -\ \frac{x dx + y dy}{r}\rt]^2 .
\ee
The coordinates $u,r,\varphi$ in $(4)$ and $t,x,y$ in $(12)$ are related
through
\bea
u\ &=&\ t\ -\ r, \nn\\
x\ &=&\ r\  \cos\varphi, \nn\\
y\ &=&\ r\  \sin\varphi.
\eea
The determinant $g \equiv \ \vline \  g_{ij}\ \vline$ is
\be
g\ = \ -1
\ee
and the contravariant components of the metric tensor
are
\bea
g^{00} &=& -m -2,\nn\\
g^{11} &=& \frac{- m x^2 + y^2}{r^2},\nn\\
g^{22} &=& \frac{- m y^2 + x^2}{r^2},\nn\\
g^{01} &=& - \ \frac{x(1+m)}{r},\nn\\
g^{02} &=& - \ \frac{y(1+m)}{r},\nn\\
g^{12} &=& - \ \frac{x y(1+m)}{r^2}\ .
\eea
We are interested in calculating the energy and momentum density components
and therefore the required components of $S^{mjnk}$ are
\bea
S^{0101}\ &=&\ \frac{-y^2(1+m)-r^2}{r^2},\nn\\
S^{0202}\ &=&\ \frac{-x^2(1+m)-r^2}{r^2},\nn\\
S^{0102}\ &=&\ \frac{x y (1+m)}{r^2},\nn\\
S^{0112}\ &=&\ \frac{y (1+m)}{r},\nn\\
S^{0212}\ &=&\ \frac{- x (1+m)}{r}\  .
\eea
Substituting $(16)$ in $(9)$, we get the energy and momentum (energy current)
density components
\bea
L^{00}\ &=& \ \frac{m \prime }{2\  \kappa\ r}  , \nn\\
L^{01}\ &=& \ - \frac{\dot{m} x}{2\ \kappa\ r^2} , \nn\\
L^{02}\ &=& \ - \frac{\dot{m} y}{2\ \kappa \ r^2}
\eea
($m \prime$ stands for $\frac{\partial m} {\partial r}$).
Therefore, the energy and momentum components are
\bea
E &=& \ m \ \frac{\pi}{\kappa},\nn\\
P_x &=& P_y = 0,
\eea
and the power output is
\be
W\ = \ -\  \frac{\pi}{\kappa}\ \ \frac{dm}{du}\ .
\ee
For the Vaidya metric, Lindquist, Schwartz, and Misner found $ E = M, P_x
= P_y = P_z = 0,$ and $W = - \frac{dM}{du}$, where $M$ is the
mass parameter in the Vaidya metric. Thus, we get similar results in its $2+1$
dimensional counterpart ($\kappa$ is not known due to the lack of a Newtonian
limit).

The exact solutions of the Einstein-massless scalar equations in $3+1$
spacetime dimensions are known in the literature $[11]$.
It is of interest to obtain a static and circularly symmetric exact solution
of these equations in $2+1$ dimensions.
The Einstein-massless scalar field equations are $[11]$
\be
R_{ij}\ -\ \frac{1}{2}\ R \ g_{ij}\ =\ \kappa \ S_{ij}\ ,
\ee
where $S_{ij}$, the energy-momentum tensor of the massless scalar field, is
given by
\be
S_{ij}\ =\ \Phi_{,i}\ \Phi_{,j}\ -\ \frac{1}{2}\ g_{ij}\ g^{ab}\ \Phi_{,a}\
          \Phi_{,b}\ ,
\ee
and
\be
\Phi_{,i}^{\ ;i}\ =\ 0 \ .
\ee
$\Phi$ stands for the massless scalar field.

A static and circularly symmetric exact solution of the above equations is
given by the line element
\be
ds^2\ =\ - B\ dt^2 + B\ dr^2 + r^2\ d\varphi^2
\ee
with
\be
B\ =\ (1-q)\ R^q,
\ee
and the scalar field
\be
\Phi\ =\ \sqrt{\frac{q}{\kappa}}\ \ \ln R,
\ee
where
\be
R\ = \ \frac{r}{r_0}\ .
\ee
$q$ stands for the scalar charge. $q = 0$ gives the flat spacetime
in $2+1$ dimensions.
The nonvanishing components of the Einstein tensor and the energy-momentum
tensor of the massless scalar field are given by
\be
G^1_1 = - G^2_ 2 = - G^0_0 = \kappa S^1_ 1 = - \kappa S^2_2 = - \kappa S^0_0
=   \frac{q}{2 \ (1-q)\ R^q \ r^2}\ \ .
\ee
The surviving independent components of the Riemann tensor are
\bea
R_{1212}\ &=&\ R_{2020}\ =\ \frac{q}{2},\nn\\
R_{1010}\ &=&\ \frac{q (q-1) R^q}{2 r^2}\ ,
\eea
and the Kretschmann invariant is
\be
R^{abcd}\ R_{abcd}\ =\ \frac{ 3 q^2}{(1-q)^2\ R^{2q}\ r^4}\ \ .
\ee
The Kretschmann invariant diverges at $r=0$  showing a curvature
singulariy there and it vanishes at infinity. However, the scalar
field diverges at $r =0$ as well as at   infinity.
The details of this Letter will be given elsewhere. Upon the completion of this
work a preprint $[12]$ came to our notice which has the nonstatic solution
given
here.
\begin{flushleft}
{\bf Acknowledgements}
\end{flushleft}
This work was  supported in part  by a Basque Government post-doctoral
fellowship. Thanks are due to A. Ach\'{u}carro, I. Egusquiza, P. S. Joshi,
and M. A. Valle for discussions.
\newpage
\begin{flushleft}

{\bf References}\\
$[1]$ Unruh W G\ 1994\ Inter. J. Mod. Phys. D {\bf 3}\ 131\\
 \ \ \ \  \  Witten E\ 1988\ Nucl. Phys.B \ {\bf 311}\ 46 \\
$[2]$ Gott J R and Alpert M\ 1984\ Gen. Rel. Grav.\ {\bf 16}\ 243\\
$[3]$ Deser S and Mazur P O \ 1985\ Class. Quantum Grav.\ {\bf 2}\ L51\\
$[4]$ Papapetrou A \ 1947\ Proc. R. Irish Acad. \ {\bf 51}\ 91\\
 \ \ \ \ \ Majumdar S D \ 1947\ Phys. Rev.\ {\bf 72}\ 390 \\
$[5]$ Giddings S, Abbott J and Kuchar K\ 1984\ Gen. Rel. Grav.\ {\bf 16}
751\\
$[6]$ Ba\~{n}ados M, Teitelboim C and  Zanelli J\ 1992\ Phys. Rev. Lett.\ {\bf
69}
\ 1849\\
$[7]$ Cangemi D, Leblanc M and Mann R B\ 1993\ Phys. Rev. D\ {\bf 48} 3606\\
 \ \ \ \  \ Kaloper N \ 1993\ Phys. Rev. D {\bf 48} \ 2598\\
 \ \ \ \  \ Ach\'{u}carro A and Ortiz M\ 1993\ Phys. Rev. D \ {\bf 48}\ 3600\\
 \ \ \ \  \ Shiraishi K and Maki T\ 1994\ Class. Quantum Grav.\ {\bf 11}\ 695\\
 \ \ \ \  \ Steif A R\ 1994\ Phys. Rev. D \ {\bf 49}\ R585\\
 \ \ \ \  \ Lifschytz G and Ortiz M\ 1994\ Phys. Rev. D \ {\bf 49}\ 1929\\
 \ \ \ \  \ Hyun S, Lee G H and Yee J H\ 1994\ Phys. Lett. B \ {\bf 322}\ 182\\
 \ \ \ \  \ Zaslavskii O B\ 1994\ Class. Quantum. Grav.\ {\bf 11}\ L33\\
$[8]$ Ba\~{n}ados M, Henneaux M, Teitelboim C and Zanelli J\ 1993\ Phys. Rev. D
\ {\bf 48}\ 1506\\
$[9]$  Lindquist R W , Schwartz R A and  Misner C W\ 1965\ Phys. Rev.\
{\bf137}\  B1364.\\
$[10]$  Landau L D and  Lifshitz E M\ 1985\  The Classical Theory of Fields (
Oxford\\
\ \ \ \ \ : Pergamon Press) p.280.\\
$[11]$ Janis A I, Robinson D C and Winicour J\ 1969\ Phys. Rev.\ {\bf 186}\
1729\\
$[12]$ Chan J S F, Chan K C K and Mann R B\ 1994\ gr-qc/9406049
\end{flushleft}
\end{document}